\input epsf
\font\upright=cmu10 scaled\magstep1
\font\sans=cmss10
\newcommand{\be}{\begin{equation}}
\newcommand{\ee}{\end{equation}}
\newcommand{\bea}{\begin{eqnarray}}
\newcommand{\eea}{\end{eqnarray}}
\newcommand{\ssf}{\sans}
\newcommand{\Z}{\hbox{\upright\rlap{\ssf Z}\kern 2.7pt {\ssf Z}}}
\newcommand{\R}{\hbox{\upright\rlap{I}\kern 1.7pt R}}
\newcommand{\PP}{\hbox{\upright\rlap{I}\kern 1.8pt P}}
\newcommand{\bn}{{\bf n}}

\newcommand{\bA}{{\bf A}}
\newcommand{\bF}{{\bf F}}
\newcommand{\bL}{{\bf L}}
\newcommand{\vn}{{\vec{n}}}
\newcommand{\vx}{{\vec{x}}}
\newcommand{\vtau}{{\vec{\tau}}}
\newcommand{\p}{{\partial}}
\newcommand{\tr}{{\rm tr}}
\newcommand{\QH}{Q_{\rm H}}
\newcommand{\QB}{Q_{\rm B}}

\documentstyle[12pt]{article}

\newcommand{\news}{\setcounter{equation}{0}}
\begin{document}
\title{
\vskip 30pt
{\bf \LARGE \bf Faddeev-Skyrme Model and Rational Maps}} 
\vskip 3.0 cm

\author{Wang-Chang Su \\[10pt]
\\{\normalsize{\sl Department of Physics}}
\\{\normalsize{\sl National Chung-Cheng University}}
\\{\normalsize{\sl Chia-Yi, 621, Taiwan}}
\\
\\{\normalsize{\sl Email : suw@phy.ccu.edu.tw}}\\ 
}
\maketitle
\vskip 25pt

\begin{abstract}

The Faddeev-Skyrme model, a modified $O(3)$ nonlinear sigma model in three space dimensions, is known to admit topological solitons that are stabilized by the Hopf charge.
The Faddeev-Skyrme model is also related to the low-energy limits of $SU(2)$ Yang-Mills theory.
Here, the model is reformulated into its gauge-equivalent expression, which turns out to be Skyrme-like. 
The solitonic solutions of this Skyrme-like model are analyzed by the rational map ansatz. 
The energy function and the Bogomolny-type lower bound of the energy are established. 
The generalized Faddeev-Skyrme model that originates from the infrared limits of $SU(N)$ Yang-Mills theory is presented.

\end{abstract} 
\newpage

\section{Introduction}\news

Topological solitons are of great importance in understanding nonperturbative effects of quantum field theories. 
From a mathematical point of view, topological solutions to the equation of motion of a quantum field theory are possible only if there is a nontrivial mapping of space-time onto target, that is, the existence of nontrivial homotopy classes.
The nontrivialness of homotopy classes is related to the possibility of nonvanishing topological charge, or soliton number, of field configurations.
In the majority of models, the soliton number arises as a winding number between spheres of equal dimension.  
Given $SU(2)$ Yang-Mills theory as an example, the instanton number is a winding number between two 3-spheres, which counts the wrapping of the gauge function as a map from the $S^3$ space to the $S^3$ target. 
As another example, in Skyrme model \cite{skyrme} the baryon number is a winding number between two 3-spheres, in which the Skyrme field is responsible for the degree of mapping.

Nevertheless, there are exceptions.
There are models in which the soliton numbers are not classified by the winding number as before, but by those with very different topological characteristics. 
The prototype in this minority was proposed by L. D. Faddeev in the mid-seventies, the so-called Faddeev-Skyrme model \cite{faddeev}.
It is a modified version of the $O(3)$ nonlinear sigma model in 3+1 dimensions.
The solitonic structure in the model is line-like, not point-like as it is in the case of instantons and skyrmions.
The soliton number is a linking number. 
The topological characteristic is the linking property of the field configurations. 
As a result, it is the linking number that is responsible for the stability of the solitons. 
 
Explicitly, the Faddeev-Skyrme model \cite{faddeev} consists of a real three-component vector field \( \vn (\vx , t) = (n^1,n^2,n^3) \), with unit length \( \vn \cdot \vn = 1 \). 
The Lagrangian density in (3+1)-dimensions takes the form
\be
{\cal L}_{\rm FS} 
\, = \, \p_\mu \vn \cdot \p_\mu \vn
\, - \, \frac{1}{4} \left( \p_\mu \vn \times \p_\nu \vn \right) \cdot
                           \left( \p_\mu \vn \times \p_\nu \vn \right) \, .
\label{FS}
\ee
The first term in the Lagrangian (\ref{FS}) represents the action of the ordinary $O(3)$ nonlinear sigma model. 
The second term, as suggested by the Derrick's scaling theorem \cite{derrick}, is introduced to prevent an instability of field configurations.

If we concentrate ourselves on the static field configurations in the model (\ref{FS}), a topological characterization results.
To see this, we observe that, for a field configuration to have finite energy, the vector field $\vn (\vx)$ must approach a constant value at spatial infinity, i.e. \( \vn (\vx) \to \vn_0 \) as \( |\vx| \to \infty \).  
Because of the boundary condition, the Euclidean $\R^3$ space can be regarded as a compactified three dimensional sphere, an $S^3$ space.
Hence, at any fixed time a vector field \( \vn (\vx) \) defines a map from the 3-sphere space to the 2-sphere target.
Note that the spheres between the mapping do not have the same dimensions ($S^3$ to $S^2$).
Such unusual mapping falls into nontrivial homotopy classes, \( \pi_3 (S^2) = \Z \). 

The topological charge associated with each static field configuration is called the Hopf charge $\QH$.
However, the Hopf charge $\QH$ cannot be expressed locally in terms of the vector field $\vn$.
A nonlocal expression is then derivable as follows.
Define \( F_{ij} \equiv \left( \p_i \vn \times \p_j \vn \right) \cdot \vn \, \) be a rank-two antisymmetric tensor, then the closed volume two-form on the target $S^2$ is given by
\( \Omega = F_{ij} \, dx^i \wedge dx^j \). 
Because the second cohomology group of the 3-sphere is trivial, i.e., \( H^2(S^3) = 0 \), the preimage of the two-form $\Omega$ under $\vn$ on the base $S^3$ must be exact.
It implies that there exists a vector $A_i$ such that \( F_{ij} = \p_i A_j - \p_j A_i \). 
In this way, the Hopf charge coincides with the integration of the three-dimensional Chern-Simons term,
\be
\QH \, = \, \frac{1}{32\pi^2} \int d^3{x} \, \epsilon_{ijk} \, F_{ij} \, A_k \,.
\label{hopfcharge}
\ee
Since the solitons are line-like, the Hopf number $\QH$ (\ref{hopfcharge}) for a given configuration represents the linking number between two different field lines.    

The existence of the topological charge (\ref{hopfcharge}) suggests a lower bound on the energy of field configuration with non-zero Hopf charge \cite{KR,ward}. 
The argument consists of a number of parts based on the method of functional analysis.
It can be shown that the bound on the energy $E$ of a static configuration with Hopf charge $\QH$ is saturated by
\be 
E \, \ge \, c \, \vert \QH \vert^\frac{3}{4} \, ,
\label{bound} 
\ee
where $c$ is a universal constant.
Its value is given by \( c = 8 \sqrt{2} \pi^2 3^{\frac{3}{8}} \) \cite{KR} and can be improved to $32 \pi^2$ \cite{ward}.
Note that the energy bound (\ref{bound}) is not obtained from the usual Bogomolny-type argument, otherwise it would be linear in the topological charge $\QH$.

The Faddeev-Skyrme model (\ref{FS}) has been under intensively numerical studies recently \cite{FN,GH,BS}. 
The numerical results have revealed the intricate and fascinating structures of the model, especially in the soliton structures with higher Hopf charge. 
However, to gain more insight and to exhibit the complete structures, one needs analytic results of the model as well. 
The purpose of this paper is to present a qualitative understanding on the soliton solutions of the Faddeev-Skyrme model. 
The paper is organized as follows.
In section 2, from the viewpoint of Yang-Mills theory, the model is reformulated into a gauge-equivalent expression that turns out to be Skyrme-like.
In Section 3, the solitonic solutions of the Skyrme-like model are analyzed by the rational map ansatz. 
A simple energy function is obtained and the Bogomolny-type lower bound of the energy is established.
In section 4, we summary the article and present a novel class of the generalized Faddeev-Skyrme models.

\section{The viewpoint of Yang-Mills theory}\news

The presence of closed line-like solitons in the Faddeev-Skyrme model (\ref{FS}) inspires many researchers, in particular Faddeev and Niemi, to consider them as the natural candidates for describing glueballs in the infrared spectrum of a Yang-Mills theory.
The motivation comes from the qualitative picture developed by 't Hooft and Polyakov, who asserted that the ultraviolet and infrared limits of a Yang-Mills theory should represent different phases. 
In the high-energy limits, the theory describes the interaction of massless gluons and can be solved perturbatively thanks to the asymptotic freedom.
Take $SU(2)$ Yang-Mills theory as an example, the Lagrangian is
\be
{\cal L}_{\rm YM} \, = \, 
\frac{1}{g^2} \, \tr \left( \bF_{\mu\nu} \bF_{\mu\nu} \right) \, ,
\label{YM}
\ee
where the non-Abelian field strength is defined by
\( \bF_{\mu\nu} = \p_\mu \bA_\nu - \p_\nu \bA_\mu - 
i \left[ \bA_\mu , \bA_\nu \right] \).
The gauge connection is
\( \bA_\mu = A^a_\mu \tau^a \),
where $\tau^a$ for $a = 1, 2, 3$ denote the Pauli matrices.

However, the Yang-Mills theory becomes strongly coupled at the low energies.
The method of perturbation theory fails. 
Despite this, we expect that the low energy limits of the theory exhibit color confinement due to the dual Meissner effect \cite{nambu,thooft}. 
Thus, the spectrum of the low-energy theory shall possess massive solitonic flux tubes, which close themselves in stable knot-like configurations.
Can the line-like solitons in the Faddeev-Skyrme model be identified with the knot-like configurations of the Yang-Mills theory at low energies?  
Faddeev and Niemi have an answer to this question. 
It is known that an Abelian projection of the Yang-Mills theory to its maximal Abelian subgroup is responsible for the dynamics of the dual Meissner effect \cite{thooft}. They are thus led to propose an on-shell decomposition of the $SU(2)$ Yang-Mills connection $\bA_\mu$, in terms of the variables in the Abelian subgroup \cite{fad1}. 
The decomposition involves a three component unit vector \( \bn = n^a \tau^a \), an Abelian gauge field $C_\mu$, and a complex scalar $\phi = \rho + i \sigma$. 
Using these fields, the $SU(2)$ gauge connection admits the following expression of Abelian decomposition
\be
\bA_\mu \, = \, C_\mu \bn \, + \, 
             \frac{1}{i} \left[ \, \p_\mu \bn , \bn \, \right] \, + \, 
             \rho \, \p_\mu \bn \, + \, 
             \sigma \frac{1}{i} \left[ \, \p_\mu \bn , \bn \, \right] \, .
\label{bA}
\ee

The first two terms on the right-hand-side of (\ref{bA}) is called the Cho connection \cite{cho1}.
Under $SU(2)$ gauge transformations in the direction $\bn$, the Cho connection retains the full non-Abelian characteristic, while the field $C_\mu$ transforms as a $U(1)$ connection.
As a matter of fact, the fields $C_\mu$ and $\phi$ determine an Abelian Higgs multiplet. 
Now, if we substitute (\ref{bA}) to the Yang-Mills Lagrangian (\ref{YM}), we obtain a non-renormalizable decomposed theory in terms of the variables 
\( ( \bn, C_\mu, \phi ) \).
This decomposition is on-shell complete in the sense that the variations of the decomposed theory with respect to the fields \( ( \bn , C_\mu, \phi ) \), respectively, reproduce the equations of motion of the original $SU(2)$ Yang-Mills theory \cite{fad1}.

There are novel features in the decomposed theory. 
On one hand, if the vector field $\bn$ is averaged over first with \( \langle \tr \left( [ \p_\mu \bn , \p_\nu \bn ] \, \bn \right) \rangle = 0 \), the multiplet \( (C_\mu, \phi) \) transforms as the fields in the Abelian Higgs model.
On the other hand, if the fields \( (C_\mu, \phi) \) of the decomposed theory are properly integrated out with \( \langle \p_\mu \phi^* \p_\nu \phi \rangle \propto \delta_{\mu\nu} \), the resultant theory coincides with the Faddeev-Skyrme Lagrangian (\ref{FS}).
This gives a strong evidence that the Faddeev-Skyrme model (\ref{FS}) is appropriate for describing the low-energy limits of Yang-Mills theory, through the Abelian decomposition (\ref{bA}) of the gauge field. 
The derivation of the Faddeev-Skyrme model from the first principles has been reported by many authors via different methods \cite{cho2,lang,gies}.  

Since the vector field $\bn$ in the decomposition (\ref{bA}) is Lie-algebra valued, it can be identified, without loss of generality, as the conjugation of the Cartan subalgebra $\tau^3$ by a generic group element $U$ in $SU(2)$.
That is,
\be
\bn \, \equiv \, U \tau^3 U^{-1} \, .
\label{bn}
\ee
With this definition (\ref{bn}), we can replace the field $\bn$ by the scalar field $U$ in the decomposition (\ref{bA}) and recast the on-shell connection $\bA_\mu$ into an explicit expression of gauge transformation as follows,
\be
\bA_\mu \, = \, U \tilde{\bA}_\mu U^{-1} \, + \, 
\frac{1}{i} \p_\mu U U^{-1}.
\label{gaugetransform} 
\ee
The manifestly gauge equivalent connection $\tilde{\bA}_\mu$ in (\ref{gaugetransform}) has the form 
\be
\tilde{\bA}_\mu \, = \, 
\left( C_\mu -  L^3_\mu \right) \tau^3 \, + \, 
\rho \, [ \, \bL_\mu , \tau^3 \, ] \, + \, 
\frac{1}{i} \sigma \left[ \, [ \, \bL_\mu , \tau^3 \, ] , \tau^3 \right] \, ,
\label{tildebA}
\ee
where the covariant current $\bL_\mu$ is an element in the $SU(2)$ Lie algebra.
It is defined by
\be 
\bL_\mu \, \equiv \, i \, L_\mu^a \, \tau^a \, = \, U^{-1} \p_\mu U \, . 
\label{bL}
\ee
Because the field $U$ plays the role of gauge function under the gauge transformation (\ref{gaugetransform}), the current $-i \, \bL_\mu$ (\ref{bL}) behaves as a pure gauge connection, and satisfies the identity \( \p_\mu \bL_\nu - \p_\nu \bL_\mu + \left[ \bL_\mu , \bL_\nu  \right] = 0 \).

Now, if we substitute (\ref{tildebA}), instead of (\ref{bA}), to the Yang-Mills Lagrangian (\ref{YM}), we obtain another expression for the decomposed theory in terms of the fields \( ( \bL_\mu, C_\mu, \phi ) \), which is gauge equivalent to the previous one mentioned.
The Abelian multiplet \( (C_\mu, \phi) \) of this latter decomposed theory can then be properly integrated over to render a gauge equivalent version of the Faddeev-Skyrme Lagrangian (\ref{FS}). 
The resultant theory is
\be
{\tilde{\cal L}}_{\rm FS} 
\, = \, 4 \, L^a_\mu \, L^a_\mu 
\, - \, 4 \left( \epsilon^{3ab} \, L^a_\mu \, L^b_\nu \right)^2 \, ,
\label{tildeFS} 
\ee
where the color indices $a,b = 1,2$. 
Note that the Lagrangian (\ref{tildeFS}) turns out to be very much Skyrme-like.\footnote{
The Skyrme model \cite{skyrme} is a non-linear theory of pions, with the Skyrme field $U(\vx , t)$ being an $SU(2)$ valued scalar.
The Lagrangian density takes the form
\[
{\cal L}_{\rm S} 
\, = \, - \frac{1}{2} \, \tr \left( \bL_\mu \, \bL_\mu \right)
\, + \, \frac{1}{16} \, \tr \left( [ \bL_\mu , \bL_\nu ]^2 \right) \, ,
\]
where $\bL_\mu$ is the $SU(2)$ Lie-algebra valued current \( \bL_\mu = U^{-1} \p_\mu U \).
For a static Skyrme field to have finite energy, it must satisfies the boundary condition \( U \to 1 \) as \( \vert \vx \vert \to \infty \).
This defines a map from $S^3$ to $SU(2)$ with nontrivial homotopy classes, \( \pi_3 (S^3) = \Z \).
The associated topological charge, the baryon number $\QB$, of the field configuration is given by
\[
\QB \, = \,
\frac{1}{24 \pi^2} 
\int d^3{x} \, \epsilon_{ijk} \, \tr \left( \bL_i \, \bL_j \, \bL_k \right) \, . 
\]
}
Further, (\ref{tildeFS}) is invariant under a residual $U(1)$ gauge transformation of $SU(2)$, which is $U \to Uh$, where $h = \exp(i \alpha \tau^3)$.
The gauge invariant (physical) fields thus have values in the coset space 
\( SU(2)/U(1) = S^2 \).
The gauge equivalent expression (\ref{tildeFS}) has been obtained recently in the paper \cite{baal}.

In terms of the covariant current $\bL_\mu$ (\ref{bL}), the corresponding Hopf charge $\QH$ of the gauge equivalent model (\ref{tildeFS}) is given by
\be
\QH \, = \,
\frac{1}{4 \pi^2} 
\int d^3{x} \, \epsilon_{ijk} \, \epsilon^{3ab} \, L^a_i \, L^b_j \, L^3_k \, .
\label{hopfcharge2} 
\ee 
It is worth pointing out that the expression (\ref{hopfcharge2}) coincides with that of the baryon number of the Skyrme model. 
Thus, the Hopf charge of the Faddeev-Skyrme model (\ref{tildeFS}) is precisely the topological charge of the Skyrme field.
Furthermore, because the current $-i \, \bL_\mu$ is a pure gauge connection (\ref{bL}), the Hopf charge of the model is also identical to the instanton number (in the temporal gauge) of the $SU(2)$ Yang-Mills theory.

\section{Rational maps and the energy}\news

Rational maps are first introduced for the study of the BPS monopole \cite{donaldson}. It is found that each rational map arises from some monopole, and conversely, the monopole can also be constructed from the rational map \cite{jarvis}.
Using the symmetry of rational maps, one can understand the scattering of monopoles with that symmetry.  
Later on, the rational maps are brought in the scene of the Skyrme model due to the similarities between monopoles and skyrmions in symmetries of low energy solutions \cite{HMS}.
Based on the rational map ansatz, the minimal energy configurations with higher baryon number are constructed, and the low-lying vibrational modes of skyrmions are studied, as well.

A rational map is a holomorphic map between two spheres.
If the first $S^2$ has coordinate $z$ and the second one has coordinate $R$, a rational map of degree $N$ is a function \( R : S^2 \to S^2 \), where \( R(z) = p(z)/q(z) \).
Here, $p$ and $q$ are polynomials of degree at most $N$, and $p$ and $q$ have no common roots.
Because a complex coordinate on two-sphere can be stereographically projected into the conventional polar coordinates \( z = \tan (\theta/2) e^{i \phi} \), the point $z$ on the first two-sphere thus corresponds to the unit vector in the Cartesian notation as
\be
\vn_z 
\, = \, \frac{1}{1 + \vert z \vert^2}
\left( z + \bar{z} , \, i (\bar{z} - z) , \, 1 - \vert z \vert^2 \right) \, .
\label{nz}
\ee
In the same vein, using the stereographical projection, there exists an expression for the unit vector of the rational map $R(z)$ on the target $S^2$
\be
\vn_R 
\, = \, \frac{1}{1 + \vert R \vert^2}
\left( R + \bar{R} , \, i (\bar{R} - R) , \, 1 - \vert R \vert^2 \right) \, .
\label{nR}
\ee

How Skyrme fields arise from the rational maps is established in \cite{HMS,krusch} using the rational map ansatz.
Let a point in $\R^3$ space has the coordinates $(r,z)$, where $r$ is the radial distance and $z$ specifies the direction. 
A static Skyrme field $U(r,z)$, whose baryon number is denoted by $\QB$, satisfies the boundary condition $U \to 1$ as $r \rightarrow \infty$.
Therefore, the Skyrme field is a map from \( \R^3 \to S^3 \), whereas the rational map is that from \( S^2 \to S^2 \).
For these two maps to be compatible, the ansatz states that the Skyrme field is expressed in terms of a radial profile function $f(r)$ and a direction in the $SU(2)$ Lie-algebra determined by the unit vector $\vn_R$ (\ref{nR}) as follows 
\be
U(r,z) 
\, = \, 
\exp \left( \, i f(r) \, \vn_R \cdot \vtau \, \right) \, ,
\label{ansatz}
\ee
where $\vtau = ( \tau_1 , \tau_2 , \tau_3 )$ are the Pauli matrices.
The boundary condition of the Skyrme field at infinity requires that $f(\infty) = 0$.
Similarly, for the ansatz (\ref{ansatz}) to be well defined at the origin $r=0$, we take \( f(0) = \pi \).
Thus, the profile function $f(r)$ is a monotonically decreasing function.
The baryon number constructed in this way is simply given by \( \QB = N \), where $N$ is the degree of the rational map $R(z)$. 

The construction of skyrmion solutions in terms of rational maps suggests that one might understand the solutions of the Faddeev-Skyrme model using rational maps as well. 
The reasoning is rooted in the similarities between the Skyrme model and the gauge equivalent formulation of the Faddeev-Skyrme model (\ref{tildeFS}).
Observe that the Skyrme model can be consistently reduced to the model (\ref{tildeFS}), by restricting the values of the $SU(2)$ Lie-algebra current $\bL_\mu$ to those in the coset representation \( SU(2)/U(1) \).
A Faddeev-Skyrme field configuration with Hopf charge $\QH$ is then obtained by applying the Skyrme filed that has baryon number \( \QB = \QH \).
Therefore, the application of the ansatz (\ref{ansatz}) infers that the degree of the rational map $N$ is precisely the baryon number $\QB$ of the Skyrme field and also the Hopf charge $\QH$ of the Faddeev-Skyrme field.

Accordingly, it is sufficient to represent the field of the Faddeev-Skyrme model (\ref{tildeFS}) by the Skyrme field. 
Then using the rational map ansatz (\ref{ansatz}), we find that the energy function of static field configurations of the Faddeev-Skyrme model (\ref{tildeFS}) takes the form
\bea
E_{\rm FS} & = & 4 \int \,
\Bigg[ \,
f^{\, 2}_r 
\frac{4 |R|^2}{\left( 1 + |R|^2 \right)^2} 
\, + \, 
\left( 1 + 2 f^{\, 2}_r \right) 
\frac{\sin^2 \! f}{r^2} 
\left( \frac{1 + |z|^2}{1 + |R|^2} \left| \frac{dR}{dz} \right| \right)^2 
\nonumber \\
& & 
\, + \, 
\left( 1 - 2 f^{\, 2}_r \right) 
\frac{\sin^2 \! f}{r^2}
\left( \frac{1 - |R|^2}{1 + |R|^2} \right)^2  
\left( \frac{1 + |z|^2}{1 + |R|^2} \left| \frac{dR}{dz} \right| \right)^2 
\nonumber \\
& &
\, + \, 
2 \, \frac{\sin^4 \! f}{r^4}
\left( \frac{1 - |R|^2}{1 + |R|^2} \right)^2
\left( \frac{1 + |z|^2}{1 + |R|^2} \left| \frac{dR}{dz} \right| \right)^4 
\Bigg] \, 
\frac{2i \, dz \, d\bar{z} \, r^2 dr}{\left( 1 + |z|^2 \right)^2} \, ,
\label{FSenergy}
\eea
where \( f_r = df/dr \) and 
\( 2i \, dz \, d\bar{z}/ {\left( 1 + |z|^2 \right)^2} = 
\sin \! \theta \, d\theta \, d\phi \) is the area element on the space two-sphere.
In the expression (\ref{FSenergy}), the part of angular integration of the second integrand is exactly the pullback of the area integral on the target two-sphere. That is,  
\be
\int \, 
\left( \frac{1 + |z|^2}{1 + |R|^2} \left| \frac{dR}{dz} \right| \right)^2
\frac{2i \, dz \, d\bar{z}}{\left( 1 + |z|^2 \right)^2} 
\, = \,
\int \,
\frac{2i \, dR \, d\bar{R}}{\left( 1 + |R|^2 \right)^2}
\, = \, 
4 \, \pi N \, ,
\label{target1}
\ee
where $N$ is the degree of the rational map $R(z)$.
As it were, the part of angular integration of the third integrand in (\ref{FSenergy}) is one-third of the pullback area on the target,
\be
\int \, 
\left( \frac{1 - |R|^2}{1 + |R|^2} \right)^2
\left( \frac{1 + |z|^2}{1 + |R|^2} \left| \frac{dR}{dz} \right| \right)^2
\frac{2i \, dz \, d\bar{z}}{\left( 1 + |z|^2 \right)^2} 
\, = \, 
\frac{1}{3} \, \left( 4\pi N \right) \, .
\label{target2}
\ee
The result (\ref{target2}) is deduced by noticing that \( \left( \frac{1 - |R|^2}{1 + |R|^2} \right)^2 = n_R^{\, 3} n_R^{\, 3} \), where $n_R^{\, 3}$ is the third component of the unit vector $\vn_R$ (\ref{nR}). 
In addition, no preferred direction exists in the rational map $R(z)$.
With (\ref{target1}) and (\ref{target2}), the energy function (\ref{FSenergy}) simplifies to
\be
E_{\rm FS} \, = \,
16 \, \pi \int dr \,
\Bigg[ \>
{\cal I} \, r^2 f^{\, 2}_r  
\, + \, 
\frac{4}{3} \, N \left( 1 + f^{\, 2}_r \right) \, \sin^2 \! f 
\, + \,
2 {\cal J} \, \frac{\sin^4 \! f}{r^2} \, 
\Bigg] \, ,
\label{FSenergy2}
\ee
where ${\cal I}$ and ${\cal J}$ both are functions on the space of rational maps and are independent of the radial distance $r$. 
Explicitly, they are
\bea
{\cal I} & = &
\frac{1}{4 \pi}
\int \, 
\frac{4 |R|^2}{\left( 1 + |R|^2 \right)^2} 
\frac{2i \, dz \, d\bar{z}}{\left( 1 + |z|^2 \right)^2} \, ,
\label{calI}
\\
{\cal J} & = &
\frac{1}{4 \pi}
\int \,
\left( \frac{1 - |R|^2}{1 + |R|^2} \right)^2
\left( \frac{1 + |z|^2}{1 + |R|^2} \left| \frac{dR}{dz} \right| \right)^4  
\frac{2i \, dz \, d\bar{z}}{\left( 1 + |z|^2 \right)^2} \, .
\label{calJ}
\eea

The simple expression in energy (\ref{FSenergy2}) signifies one of the advantages of using the rational map ansatz (\ref{ansatz}).
To obtain the minimal energy solution, one can follow the standard procedure, which is to minimize the energy (\ref{FSenergy2}) with respect to the rational map $R(z)$ for a given degree $N$, and to the profile function $f(r)$.
Since there are two rational map functions ${\cal I}$ and ${\cal J}$, one has to {\sl simultaneously} minimize both functions using the same map over the space of rational maps of degree $N$.
Then, the profile function $f(r)$ is found by solving a second order Euler-Lagrangian equation with $N$, ${\cal I}$, and ${\cal J}$ treated as constants.
However, if the minimization for both ${\cal I}$ and ${\cal J}$ cannot be reached by the same rational map, one should consider the set \( ({\cal I},{\cal J}) \) as a two-dimensional parameter space.
Then, one finds the particular point(s) \( ({\cal I}_0,{\cal J}_0) \) in the parameter space such that the globally minimal energy of (\ref{FSenergy2}) is achieved through the variation of the profile function $f(r)$.
In practice, one learns from the study of the Skyrme model that the solutions obtained in this way are in general not exact solutions to the equation of motion, but close approximations \cite{HMS,BS}.
Nevertheless, they do provide sufficient information for the search of exact solutions.
Since the calculation mentioned above has to be done numerically, we shall not pursue this direction here, but leave it for further investigation.
 
The other advantage of using the rational map ansatz (\ref{ansatz}) is that a Bogomolny-type low-bound of the energy emerges naturally.
To see this, let us complete the square in the energy expression (\ref{FSenergy2}) by a Bogomolny-type argument, 
\bea
E_{\rm FS} 
& = &
16 \, \pi \int dr \,
\Bigg[ \,
\left( \, \sqrt{{\cal I}} \, r f_r + \sqrt{2 {\cal J}} \, \frac{\sin^2 \! f}{r}  \, \right)^2
\, + \, 
\frac{4}{3} \, N \left( 1 + f_r \right)^2 \, \sin^2 \! f \,
\Bigg]
\nonumber \\
& &
\, - \, 
32 \, \pi 
\left( \sqrt{2 \, {\cal I} {\cal J}} \, + \, \frac{4}{3} \, N \right)
\int dr \, f_r \, \sin^2 \! f \, .
\label{bogonolmy}
\eea
The second integral in (\ref{bogonolmy}) can be evaluated using the boundary values of the profile function, i.e., \( f(0) = \pi \) and \( f(\infty) = 0 \).
Because the first integral in (\ref{bogonolmy}) is always positive, the energy is thus bounded from below by the second integral alone
\be
E_{\rm FS} \, \ge \,
16 \, \pi^2 
\left( \, \frac{4}{3} N \, + \, \sqrt{2 \, {\cal I} {\cal J}} \, \right)
\, \ge \, \frac{64}{3} \, \pi^2 N \, .
\label{bogonolmy2}
\ee
The last inequality of (\ref{bogonolmy2}) uses the elementary Schwartz inequality
\bea
{\cal I} {\cal J}
& = &
\frac{1}{\left( 4 \pi \right)^2}
\left(
\int \, 
\frac{4 |R|^2}{\left( 1 + |R|^2 \right)^2} \> dS
\right) 
\left(
\int \,
\left( \frac{1 - |R|^2}{1 + |R|^2} \right)^2
\left( \frac{1 + |z|^2}{1 + |R|^2} \left| \frac{dR}{dz} \right| \right)^4 dS
\right)
\nonumber \\
& \ge &
\frac{1}{\left( 4 \pi \right)^2}
\left(
\int \,
\frac{2 |R| \left( 1 - |R|^2 \right)}{\left( 1 + |R|^2 \right)^2} \,
\frac{2i \, dR \, d\bar{R}}{\left( 1 + |R|^2 \right)^2}
\right)^2 \, = \, 0 \, ,
\eea
where \( dS \equiv 2i \, dz \, d\bar{z}/\left( 1 + |z|^2 \right)^2 \).

The rational map bound (\ref{bogonolmy2}) is valid, if the Faddeev-Skyrme field obeys the rational map ansatz.
It implies that the winding number of the rational map $N$ is precisely the Hopf charge of the field configuration $\QH$. 
Hence, the Bogonolmy-type of bound becomes \( E_{\rm FS} \ge \frac{64}{3} \pi^2 \QH \).
When comparing this linear bound with the fractional bound (\ref{bound}), we find that the former bound is weaker than the latter bound for configurations with large Hopf charge.
However, these two bounds are identical in the case of Hopf charge unity.   
As a result, the value of the universal constant $c$ given by (\ref{bound}) can be further improved to \( c = \frac{64}{3} \pi^2 \), when the rational map ansatz is applied.

The rational map ansatz leads to the simple expression of energy (\ref{FSenergy2}), with two functions ${\cal I}$ and ${\cal J}$ on the space of rational maps of any given degree $N$.
We do not attempt to present the numerical results of these two rational functions, instead we restrict to discuss the case of field configurations with Hopf charge $\QH=1$.
The corresponding rational map function is simply the choice $R(z) = z$. 
It gives rise to the spherically symmetric Skyrme field (\ref{ansatz}), but this symmetry is broken down to an axial symmetry because of Abelian gauge symmetry of the Lagrangian (\ref{tildeFS}). 
Hence, the most symmetric structure that a Faddeev-Skyrme soliton can have is axially symmetric.

For the choice $R(z) = z$, we find the values of the rational functions \( {\cal I} = \frac{2}{3} \) and \( {\cal J} = \frac{1}{3} \).
The energy expression (\ref{FSenergy2}) in this case becomes
\be
E_{\rm FS} \left( \QH = 1 \right) \, = \,
\frac{32}{3} \, \pi \int dr \,
\Bigg[ \>
r^2 f^{\, 2}_r  
\, + \, 
2 \, \left( 1 + f^{\, 2}_r \right) \, \sin^2 \! f 
\, + \,
\frac{\sin^4 \! f}{r^2} \, 
\Bigg] \, .
\label{QH1}
\ee
Observe that, up to an over-all normalization constant, the energy expression (\ref{QH1}) of Hopf charge $\QH=1$ is exactly equivalent to that of the standard hedgehog solution with the usual profile function in the Skyrme model \cite{HMS}.
Therefore, diversified methods of estimating the minimum energy of (\ref{QH1}) are available, for examples, by using variational, relaxation, or other numerical methods.
The result is found to be \( 1.232 \times 32 \pi^2 \).
This is consistent with the Bogonolmy bound of energy (\ref{bogonolmy2}) and is also comparable with the numerical investigations \cite{FN,GH,BS}

\section{Conclusion}\news

It is known that the Faddeev-Skyrme model (\ref{FS}) can be related to the low-energy limits of $SU(2)$ Yang-Mills theory, based on the method of Abelian decomposition of gauge fields.
In this paper, the Faddeev-Skyrme model is reformulated into its gauge-equivalent expression, which turns out to be Skyrme-like (\ref{tildeFS}).
A rational map ansatz is then introduced for the investigation of this Skyrme-like model due to the similarity with the original Skyrme model.
This allows us to construct the simple form of energy function (\ref{FSenergy2}) and to establish the Bogonolmy-type of energy lower bound (\ref{bogonolmy2}).
For Hopf charge $\QH=1$, the energy function gives a good agreement with the result of the numerical studies.  
Clearly, the energy function for solitons with higher Hopf charge is desirable to be investigated in detail and this work is currently underway. 

To conclude the paper, we present a novel class of generalized Faddeev-Skyrme models, which might possess interesting topological characteristics. 
Note that the attractive feature of the gauge equivalent Lagrangian (\ref{tildeFS}) is not unique to the $SU(2)$ Yang-Mills theory.
It can be generalized to other gauge groups as well, based on the method of Abelian decomposition of gauge field for higher rank group \cite{fad2,su}.
As an example, we take $SU(N)$ Lie group whose dimension is $N^2-1$ and rank is $N-1$.
The generalization to other Lie groups is straightforward.   
After Abelian decomposing the gauge field and then integrating out all of the irrelevant variables, we obtain the infrared limits of the $SU(N)$ Yang-Mills theory as follows
\be
{\cal L}
\, = \, 4 \, L^a_\mu \, L^a_\mu 
\, - \, 4 \left( f^{\, iab} \, L^a_\mu \, L^b_\nu \right)^2 \, ,
\label{sun} 
\ee
where the indices $a,b$ specify those of the coset generators in \( SU(N)/U(1)^{N-1} \), the index $i$ denotes that of the generators in \( U(1)^{N-1} \) Cartan subalgebra, and $f^{\, iab}$ is the structure constants of the group $SU(N)$.
The covariant current is \( \bL_\mu = U^{-1} \p_\mu U \), where $U$ is a generic element in the $SU(N)$ group. 
Observe that the model (\ref{sun}) very resembles the $SU(N)$ Skyrme model.
Since \( \pi_3 ( SU(N)/U(1)^{N-1} ) = \Z \), for static field configurations the model (\ref{sun}) is expected to have solitons with nontrivial topological charge.
Based on a similar discussion in the Faddeev-Skyrme model, the soliton solutions might be analyzed by applying the associated Skyrme fields of the $SU(N)$ Skyrme model.
It would be interesting to understand the detailed structure of this model (\ref{sun}).

\section*{Acknowledgment}

\noindent 
The author is grateful to C. R. Lee for useful discussions. 
This work was supported in part by Taiwan's National Science Council Grant No. 89-2112-M-194-022.

\end{document}